\documentclass[9pt]{article}
\usepackage{amsmath}
\usepackage{amssymb}
\usepackage{amsthm} 
\usepackage[a4paper, total={13.5cm, 25cm}]{geometry}
\usepackage{graphicx}
\usepackage{epstopdf}
\usepackage{fancyhdr}
\usepackage{mathtools}
\usepackage{lastpage}
\usepackage{lineno} 
\usepackage{pdflscape} 
\usepackage[square,sort,comma,numbers]{natbib}
\usepackage{hyperref}
\usepackage{tikz}

\allowdisplaybreaks 

\newtheorem{proposition}{\bf Proposition}
\newtheorem{lemma}{\bf Lemma}

\newcommand\eqv[1]{\sim_{#1}}
\newcommand\WH{\mathbf{W_H}}

\newlength{\depthofsumsign}
\begin{document}


\title{Evolutionary graph theory revisited: general dynamics and the Moran process\\
{\Large Karan Pattni\footnote{Department of Mathematics, City University London, Northampton Square, London, EC1V~0HB, UK}
, Mark Broom$^*$, Jan Rycht\'a\v r\footnotemark[2] and \mbox{Lara J. Silvers$^*$}
}
}
	
\maketitle

\footnotetext[2]{
Department of Mathematics and Statistics, The University of North Carolina at Greensboro, Greensboro, NC 27412, USA
}

\begin{abstract}
Evolution in finite populations is often modelled using the classical Moran process. 
Over the last ten years this methodology has been extended to structured populations using evolutionary graph theory. 
An important question in any such population, is whether a rare mutant has a higher or lower chance of fixating (the fixation probability) than the Moran probability, i.e. that from the original Moran model, which represents an  unstructured population. 
As evolutionary graph theory has developed, different ways of considering the interactions between individuals through a graph and an associated matrix of weights have been considered, as have a number of important dynamics. 
In this paper we revisit the original paper on evolutionary graph theory in light of these extensions to consider these developments in an integrated way. 
In particular we find general criteria for when an evolutionary graph with general weights satisfies the Moran probability for the set of six common evolutionary dynamics.
\end{abstract}

\section{Introduction}
When modelling population evolution we are concerned with the spread of heritable characteristics in successive generations.
The type of model that is used depends upon whether the population size is assumed to be finite or infinite.
The majority of classical evolutionary models (see for example \cite{1982Smith,hofbauer1998evolutionary}) use infinite populations, although finite population models are also well estiablished, the most important models being those in \cite{1958Moran, moran1962statistical}.
These models are stochastic, and are solved using classical Markov chain methodology \citep{1975KarlinTaylor, landauer1987diffusive, AntalScheuring2006}.
See also \cite{taylor2004evolutionary, nowak2004emergence} for an extension to evolutionary games in finite populations.

The populations in the models described above, however, were ``well-mixed'', i.e. every individual was equally likely to encounter every other individual. 
Real populations of course contain structural elements, such as geographical location or social relationship, which mean that some pairs individuals are more likely to interact than others. 
In such circumstances we need to be able to identify distinct individuals (or at least distinct classes of individuals), and considering finite populations is perhaps more natural than infinite ones (although finite structures each containing an infinite number of individuals, so called ``island models'', were considered in \cite{1970Maruyama}).
In \cite{2005LiebermanEtal} the modelling ideas of \cite{1958Moran} were extended to consider such structured populations based upon graphs, known as evolutionary graph theory. This has proved very successful, spawning a large number of papers (for example \cite{allen2012mutation,2006AntalRednerSood,2008BroomRychtar, broom2011evolutionary,2006OhtsukiEtal,2007OhtsukiEtal,shakarian2012fast,voorhees2013fixation}). For informative reviews see \cite{AllenNowak2014, 2012ShakarianRoosJohnson}.

In an evolving population, we need to consider the mechanism of how the population changes, called the dynamics. 
Informally, the dynamics specify the way in which heritable characteristics are passed on from one generation to the next. 
For infinite populations the classical replicator equation \cite{taylor1978evolutionary} is often used (although there are a number of alternatives), and in the stochastic model of \cite{1958Moran} there is a natural replacement dynamics built in. 
For structured populations this issue is actually considerably more complex, and the order of births and deaths, and where selection acts, is of vital importance \citep{2009_MasudaOhtsuki,2011HadjichrysanthouEtal}. 
We shall consider a set of dynamics that are commonly used in evolutionary graph theory models. 
The relationship between dynamics and structure is of key interest because the spread of heritable characteristics is directly dependent upon it. 
Whilst having essentially no effect on populations with no structure, this relationship potentially yields very different results on graphs.

Under some circumstances it is, however, possible for the dynamics and structure to interact in such a way that the spread of heritable characteristics behaves just as if the population was homogeneous.
This was a central theme of the classic paper \cite{2005LiebermanEtal}, where two important results, the circulation theorem and the isothermal theorem, were developed that addressed this question (see also \cite{1981Slatkin} for related work).
In this paper we generalise the work of \cite{2005LiebermanEtal} to obtain a complete classification of when the combination of a population structure and dynamics can be regarded as equivalent to a homgeneous population in a precisely defined way, for the six most common evolutionary dynamics and graphs with general weights.

\begin{table}[h!]
\begin{tabular}{clp{8cm}llll}
\multicolumn{3}{c}{\large \textbf{Summary of Notation}}
\\
\hline
\vspace{-0.3cm}
\\
\textit{Symbol}
& \textit{Definition}
& \textit{Description}
\\
$N$
& $\in \mathbb{Z}^+\setminus \{0,1\}$
& Population size.
\\
$A,B$
&
& The two types of individuals in population.
\\
$I_n$
&
& Individual $n$.
\\
$S$
& $=\{n : I_n \text{ of type } A\}$
& State of the population.
\\
$\mathcal{N}$
& $=\{1,2,\ldots,N\}$
& State in which all $I_n$ of type $A$.
\\
$r$
& $\in(0,\infty)$
& Fitness of a type $A$ individual.
\\
$F_n(S)$
& $ \in \{1,r\}$
& Fitness of $I_n$ in state $S$.
\\
$D$
& $= (V,E)$
& Replacement digraph with vertices $V$ where $|V|=N$ and directed edges $E$.
\\
$w_{ij}$
& $\in [0,\infty)$
& Edge weight such that $w_{ij}>0$ if and only if \mbox{$(i,j)\in E$}.
\\
$\mathbf{W}$
& $=(w_{ij})$
& \emph{Replacement matrix:} $N\times N$ weighted adjacency matrix of tuple $(D,w)$.
\\
$T^+_n$
& $=\sum_{j=1}^N w_{nj}$
& \emph{Out temperature:} Sum of all outgoing edge weights of vertex $n \in V$.
\\
$T^-_n$
& $=\sum_{i=1}^N w_{in}$
& \emph{In temperature}: Sum of all incoming edge weights of vertex $n \in V$.
\\
$b_i$
& $\in [0,1]$
& Probability $I_i$ chosen for birth.
\\
$d_{ij}$
& $\in [0,1]$
& Probability a copy of $I_i$ replaces $I_j$ given $I_i$ was chosen for birth, i.e.\ replacement by death.
\\
$d_j$
& $\in[0,1]$
& Probability $I_j$ chosen for death.
\\
$b_{ij}$
& $\in [0,1]$
& Probability a copy of $I_i$ replaces $I_j$ given $I_j$ is chosen for death, i.e.\ replacement by birth.
\\
$\mathfrak{r}_{ij}$
& $\in [0,1]$
& Probability a copy of $I_i$ replaces $I_j$.
\\
$P_{SS'}$
& $\in [0,1]$
& State transition probability.
\\
$\mathbf{S}$
& $= (P_{SS'})$
& State transition matrix.
\\
$\mathcal{E}_{*,\mathbf{W},r}$
&
& Stochastic process with state transition matrix $\mathbf{S}$ such that $*$ dynamics are used on graph $\mathbf{W}$ and type $A$ individuals have fitness $r$.
\\
$ \rho_S^A$
& $\in [0,1]$
& Fixation probability of type $A$ individual given initial state
$S$.
\\
	$W$
&
&	Set of all strongly connected replacement matrices.
\\
	$W_C$
&	$ \{\mathbf{W}:T^+_n=T^-_n \ \forall n \}$
&	Replacement matrices that are circulations.
\\
	$W_I$
&	$ \{\mathbf{W}:T^+_i=T^-_j \ \forall i,j \}$
&	Replacement matrices that are isothermal.
\\
	$W_R$
& 	$ \{\mathbf{W}:T^+_n=1 \ \forall n\}$
&  	Right stochastic replacement matrices.
\\
	$W_L$
& 	$ \{\mathbf{W}:T^-_n=1 \ \forall n\}$
&  	Left stochastic replacement matrices.	
\\
	$C_N$
& 	
& 	Replacement matrices whose digraphs are cycles of length $N$.
\\
	$f_R$
&	$(w_{ij}) \mapsto {(w_{ij}}/{\sum_nw_{in}})$
&	Map from $W$ to $W_R$.	
\\
	$f_L$
&	$(w_{ij}) \mapsto {(w_{ij}}/{\sum_nw_{nj}})$
& 	Map from $W$ to $W_L$.
\\
	$f'$
&	$(w_{ij}) \mapsto {(w_{ij}}/{\sum_{n,k}w_{nk}})$
&	Map from $W$ to $W$.
\\
	$M_*$
& 	
& 	Replacement matrices for which $\mathcal{E}_*$ is $\rho$-equivalent to a Moran process when $*$ dynamics are used.
\end{tabular}
\caption{Notation used in this paper.}
\label{Notation}
\end{table} 

\section{The Model}
We shall first describe the population model of \cite{2005LiebermanEtal}, which generalises the model of \cite{1958Moran} by incorporating a replacement structure.
The notation used in this paper is summarised in \mbox{Table \ref{Notation}}.\\
\emph{The population has a constant size  $N\in \mathbb{Z}$, $N\ge 2$, consisting of individuals $I_1, \ldots, I_N$.
Every individual is either of type $A$ or $B$.}\\
\indent
This implies that there are $2^N$ different states of the population given by the combination of type $A$ and $B$ individuals.
We represent each state by a set $S$ such that $n\in S$ if an individual $I_n$ is of type $A$.
We can easily revert to using the number of type $A$ individuals, $|S|$, if the population is homogeneous.
The states $\emptyset$ and $\mathcal{N}=\{1,2,\ldots,N\}$ have only type $B$ and $A$ individuals respectively.\\
\emph{Individuals have a constant fitness that may depend upon their type.}\\
\indent
The fitness of individuals in state $S$ is thus given by the vector $\mathbf{F}(S)=\left(F_{n}(S)\right)_{n=1,2,\ldots,N}$ where
\begin{linenomath*}
\begin{align*}
	F_n(S)=
	\begin{cases}
		1 & n \notin S,\\
		r \in (0,\infty) & n \in S,
	\end{cases}
\end{align*}
\end{linenomath*}
is the fitness of $I_n$.
Here the fitness $r$ of a type $A$ individual is given relative to the fitness of a type $B$ individual assumed to be 1.\\
\emph{During a stochastic replacement event (that happens in an instant) an exact copy of an individual $I_i$ replaces an individual $I_j$.}\\
\indent
The replacement events may be restricted in the sense that not all individuals can replace one another.
To enforce such restrictions, \cite{2005LiebermanEtal} imposed a replacement structure using a weighted directed graph given by the tuple $(D,w)$ 
where $D=(V,E)$ is a directed graph, with sets $V$ of vertices and $E$ of directed edges, and $w$ is a map that assigns a weight to each edge such that $w:V\times V\to [0,\infty):(i,j)\mapsto w_{ij}$.
Each vertex $n\in V$ represents $I_n$ therefore $V=\{1,2,\ldots,N\}$ so $|V|=N$.
We assume that $(i,j)\in E$ if and only if $w_{ij}>0$, which indicates that $I_i$ can replace $I_j$.
Note that we allow $w_{ii}>0$ and therefore $I_i$ can replace itself.
All the information contained within the weighted digraph $(D,w)$ is conveniently summarised by the $N\times N$ weighted adjacency matrix $\mathbf{W}=(w_{ij})$ and therefore we will refer to $(D,w)$ using $\mathbf{W}$, which we call the \emph{replacement matrix}.

The replacement events are stochastic which means that there is a probability $\mathfrak{r}_{ij} = \mathfrak{r}_{ij}(\mathbf{F}(S),\mathbf{W})$ associated
with (a copy of) $I_i$ replacing $I_j$.
There are several potential \emph{evolutionary dynamics on graphs} that govern how the probability is determined.
There three main types of dynamics  that are summarised below, see also \cite{2012ShakarianRoosJohnson}. We use the convention that $I_i$ is chosen for birth and $I_j$ is chosen for death.
\begin{enumerate}
	\item \emph{Birth-Death} (BD):
	$I_i$ is chosen first then $I_j$.
	We have that $i \in V$ is chosen with probability $b_i$ and then $(i,j) \in E_{i}$ is chosen with probability $d_{ij}$, where $E_{i}$ are all edges starting in vertex $i$.
	$d_{ij}$ is used to signify that there is `replacement by death'.
	Finally, $\mathfrak{r}_{ij}=b_i d_{ij}$.
	\item \emph{Death-Birth} (DB):
	$I_j$ is chosen first then $I_i$.
	We have that $j\in V$ is chosen with probability $d_j$ and then $(i,j) \in E_{ j}$ is chosen with probability $b_{ij}$, where $E_{ j}$ are all edges ending in vertex $j$.
	$b_{ij}$ is used to signify that there is `replacement by birth'.
	Finally, $\mathfrak{r}_{ij}=d_i b_{ij}$.
	\item \emph{Link} (L):
	$I_i$ and $I_j$ are chosen simultaneously.
	In this case $(i,j)\in E$ is simply chosen with probability $\mathfrak{r}_{ij}$.
\end{enumerate}
For each type of these dynamics, the natural selection can, through the fitness parameter, influence either the choice at birth (resulting in adding ``B'') or at death (adding ``D'').
It yields 6 kinds of evolutionary dynamics on graphs summarized in
 Table \ref{StandardDynamics}. These dynamics have been extensively studied, in particular, see \cite{2009Masuda} for a detailed comparison of them.
Of these, the BDB and LB dynamics were used in \cite{2005LiebermanEtal}.

\newcommand\ds{\displaystyle}
\begin{table}
\hspace*{-0.5cm}\begin{tabular}{lccccc}
\hline
\vphantom{\large $\ds \int$} 
	Process 
	& $\mathbb{P}(I_i \text{ replaces } I_j)$ 
	& Order chosen 
	& $\mathbb{P}(\text{Chosen first})$ 
	& $\mathbb{P}(\text{Chosen second})$ 
\\
    \hline
	BDB \citep{2005LiebermanEtal}
	& $\mathfrak{r}_{ij} = b_id_{ij}$ 
	& $I_i$ then $I_j$
	& $\ds b_i = \frac{F_i(S)}{\displaystyle \sum_n F_n(S)}$ 
	& $\ds d_{ij} = \frac{w_{ij}}{\ds \sum_{n} w_{in}}$ 
\\
    	BDD \citep{2009Masuda}
	& $\mathfrak{r}_{ij} = b_id_{ij}$
	& $I_i$ then $I_j$ 
	& $\ds b_i = \frac{1}{N}$
	& $\ds d_{ij} = \frac{w_{ij}/F_j(S)}{\ds \sum_n w_{in}/F_n(S)}$
  	\\
	DBD \citep{2006AntalRednerSood}
	& $\mathfrak{r}_{ij} = d_ib_{ij}$
	& $I_j$ then $I_i$
	& $\ds d_j = \frac{1/F_j(S)}{\ds \sum_n 1/F_n(S)}$
	& $\ds b_{ij} = \frac{w_{ij}}{\ds \sum_n w_{nj}}$
\\
	DBB \citep{2006OhtsukiEtal}
	& $\mathfrak{r}_{ij} = d_ib_{ij} $
	& $I_j$ then $I_i$
	& $\ds d_j = \frac{1}{N}$
	& $\ds b_{ij} = \frac{w_{ij}F_i(S)}{\ds \sum_n w_{nj}F_n(S)}$
\\
	LB \citep{2005LiebermanEtal} 
	& $\ds \mathfrak{r}_{ij} = \frac{w_{ij}F_i(S)}{\ds \sum_{n,k}w_{nk}F_n(S)}$
	& Simultaneous
	& N/A
	& N/A
\\
	LD \citep{2009_MasudaOhtsuki}
	& $\ds \mathfrak{r}_{ij} = \frac{w_{ij}/F_j(S)}{\ds \sum_{n,k}w_{nk}/F_k(S)}$ 
	& Simultaneous
	& N/A
	& N/A
\\
\vspace{-2mm}
\\
\hline
\end{tabular}\hspace*{-0.5cm}
\caption{
List of the dynamics used in this paper. 
Note that L will be used in place of LB and LD where appropriate.
}
\label{StandardDynamics}
\end{table}

\subsection{The fixation probability}
The fixation probability, $\rho_S^A=\rho_S^A(*,\mathbf{W},r)$, is the probability that the population with initial state $S$ is absorbed in $\mathcal{N}$ where $*$ is the dynamics being used.

Given that the replacement events are random, the transitions between the states of the population are described by a stochastic process, which we denote $\mathcal{E}$. 
The properties of  $\mathcal{E}$ can be investigated once the state transition probabilities of moving from state $S$ to $S'$, \mbox{$P_{SS'}=P_{SS'}(*,\mathbf{W},r)$}, are calculated using the replacement probabilities as follows:
\begin{linenomath*}
\begin{align*}
P_{SS'} =
\begin{cases}
	\displaystyle
	\sum_{i\notin S} \mathfrak{r}_{ij}(\mathbf{F}(S),\mathbf{W}) &
	\text{if } S' = S \setminus \{j\} \text{ for some } j \in S,\\
	\displaystyle
	\sum_{i\in S} \mathfrak{r}_{ij}(\mathbf{F}(S),\mathbf{W}) &
	\text{if } S' = S \cup \{j\} \text{ for some } j \notin S,\\
	\displaystyle
	\sum_{\substack{i,j \in S\\ \vee i,j \notin S}} \mathfrak{r}_{ij}(\mathbf{F}(S),\mathbf{W}) & \text{if } S' = S.
\end{cases}
\end{align*}
\end{linenomath*}
The transition probabilities, $P_{SS'}$, satisfy the Markov property because they only depend upon the state $S$, that is, the probability of transitioning from the present state to another state is independent of any past and future state of the population.
The stochastic process $\mathcal{E}_{*,\mathbf{W},r}$ with state transition matrix $\mathbf{S}=\mathbf{S}(*,\mathbf{W},r) =(P_{SS'})_{S,S'\subset\{1,2,\ldots,N\}}$ is therefore a Markov chain.
The Markov chain $\mathcal{E}_{*,\mathbf{W},r}$ is part of the class of evolutionary Markov chains described in \cite{2014AllenTarnita}.

The absorbing states of $\mathcal{E}_{*,\mathbf{W},r}$ are $\emptyset,\mathcal{N}$, which means that if the population is in either one of these states then it remains there indefinitely.
This property of $\mathcal{E}_{*,\mathbf{W},r}$ can be used to measure the success of a type $A$ individual by calculating the probability that it fixates, that is, everyone in the population is of type $A$.
The fixation probability is then given by solving
\begin{linenomath*}
\begin{align}
	\label{Fixation}
	\rho^{A}_S= \sum_{S'\subset\{1,2,\ldots,N\}} P_{SS'}\rho_{S'}^A
\end{align}
\end{linenomath*}
with boundary conditions $\rho_{\emptyset}^A = 0$ and $\rho_{\mathcal{N}}^A = 1$.

As demonstrated in \cite{2009Masuda}, LB and LD dynamics may differ in time scale but they
yield the same fixation probabilities when fitness is constant (which is our case). Thus, for our purposes the dynamics are the same  and
we will thus consider them together and denote them by L.

\subsection{The Moran Process}
The Moran process \citep{1958Moran}, a stochastic birth-death process on finite fixed homogenous population, can be reconstructed as  $\mathcal{E}_{\text{BDB},\mathbf{W_H},r}$ for a constant replacement matrix
\begin{linenomath*}
\begin{align}
\label{eq:homogeneous}
\mathbf{W_H} = (1/N)_{i,j}.
\end{align}
\end{linenomath*}
For any $r\in (0,\infty)$ and any $S\subset \{1, \ldots, N\}$,  the fixation probability for this process, or \emph{Moran probability}, is given by
\begin{linenomath*}
\begin{align}
	\rho_{S}^A=
	\begin{cases}
		\displaystyle \frac{1-r^{-|S|}}{1-r^{-N}} & \text{if } r \ne 1,\\
		\displaystyle |S|/N & \text{if } r = 1.
	\end{cases}
\nonumber
\end{align}
\end{linenomath*}
We are interested in characterizing graphs (and evolutionary dynamics) that yield the same fixation probabilities as the homogeneous matrix $\WH$ given in \eqref{eq:homogeneous}. We note that for this matrix all of the transition probabilities $\mathfrak{r}_{ij}$ take the same value independent of $i, j$ or the dynamics, and consequently the fixation probability under each of the dynamics is the same.

\subsection{Classes of Graphs/ Matrices}
The set of all admissible replacement matrices is defined as follows
\begin{linenomath*}
\begin{align}
W = \{\mathbf{W}: \text{for every $i,j$, there is $n$ such that ($\mathbf{W}^n)_{i,j} >0$}\}.
\nonumber
\end{align}	
\end{linenomath*}
This definition means  that $\mathbf{W}$ is strongly connected as for any pair of vertices $i$ and $j$, there is a path (of length $n$) going from $i$ to $j$. 
Unless specified otherwise, we will consider admissible replacement matrices only.

As in \citep{2005LiebermanEtal}, for any $\mathbf{W}$ (admissible or not) we define the \emph{in temperature} of $I_n$, $T^-_n$, and the \emph{out temperature} of $I_n$, $T^+_n$, by 
\begin{linenomath*}
\begin{align}
	T^-_n= \sum_{j =1}^N w_{jn}\quad \text{and} \qquad
    T^+_n=\sum_{j =1}^N w_{nj}.
\nonumber
\end{align}
\end{linenomath*}

$\mathbf{W}$ is called a \emph{circulation} if $T^+_n=T^-_n$, for all $n\in V$
and it is called \emph{isothermal} if $T^+_i=T^-_j$, for all $i,j\in V$.
$\mathbf{W}$ is called \emph{right} stochastic if $T_n^+ =1$, for all $n\in V$ and
it is called \emph{left} stochastic if $T_n^- =1$, for all $n\in V$.
The sets of all circulations, isothermal matrices, right stochastic matrices, and  left stochastic matrices, respectively
are denoted by $W_C, W_I, W_R,$ and $W_L$ respectively.

The set $C_N$ denotes the sets of matrices representing \emph{cycles} of length $N$, more specifically, for $(w_{ij}) \in C_N$ we have $w_{ii}=1/2$ for $i=1,2,\ldots N$, $w_{i_1i_2}=\cdots=w_{i_ni_{n+1}}=\cdots=w_{i_{N-1}i_N}=w_{i_Ni_1}=1/2$ for some permutation  $i_1,i_2,\ldots,i_N$  of the sequence $1,2,\ldots,N$, and $w_{ij}=0$ otherwise.

We also define the maps $f_R:W\to W_R$, $f_L:W\to W_L$, and $f':W\to W$ respectively, by
\begin{linenomath*}
\begin{align}
f_R\left((w_{ij})\right)=\left(\frac{w_{ij}}{\sum_{n}w_{in}}\right), \quad
f_L\left((w_{ij})\right)=\left(\frac{w_{ij}}{\sum_{n}w_{nj}}\right), \quad \text{and}\quad
f'\left((w_{ij})\right)=\left(\frac{w_{ij}}{\sum_{n,k}w_{nk}}\right).
\nonumber
\end{align}
\end{linenomath*}
Note that $f_R$ preserves right stochastic matrices and $f_L$ preserves left stochastic matrices. Moreover, $f_R(\mathbf{W}) = f_L(\mathbf{W})$ for all $\mathbf{W}\in W_I$.
Also, since $f'$ simply involves multiplying $\mathbf{W}$ by the constant $1/\sum_{n,k} w_{nk}$, it implies that
$	\mathbf{W} \in W_\text{C} \Leftrightarrow f'(\mathbf{W})\in W_\text{C}$.

When the dynamics $*$, matrices $\mathbf{W_1}$ and $\mathbf{W_2}$, and fitness $r$ are given, we say that an evolutionary Markov chain $\mathcal{E}_{*,\mathbf{W_1},r}$ is  \emph{$\rho$-equivalent} to $\mathcal{E}_{*,\mathbf{W_2},r}$
if for every $S\subset\{1,\ldots,N\}$, $\rho_S^A(*,\mathbf{W_1},r) = \rho_S^A(*,\mathbf{W_2},r)$, in which case we write
$\mathbf{W_1}\eqv{*,r} \mathbf{W_2}$.

We are specifically interested in finding matrices equivalent to the Moran process.
For a dynamics $*$, we define
\begin{linenomath*}
\begin{align*}
M_* = \{\mathbf{W}: \mathbf{W}\eqv{*,r} \WH \text{ for all } r>0 \}.
\end{align*}
\end{linenomath*}

\section{Results}
The map $f_R$ preserves the equivalence classes of BDB and BDD dynamics,
$f_L$ preserves the equivalence classes of DBB and DBD dynamics and $f'$ preserves the equivalence classes for link dynamics. Specifically, as one can see from the proofs in the Appendix, for any $\mathbf{W}$ and any $r>0$
\begin{linenomath*}
\begin{align}
\mathbf{W} &\eqv{\text{BDB},r} f_R(\mathbf{W}), \label{eq:fR equivalence of BDD}
\\
\mathbf{W} &\eqv{\text{BDD},r} f_R(\mathbf{W}),
\nonumber
\\
\mathbf{W} &\eqv{\text{DBB},r} f_L(\mathbf{W}),
\nonumber
\\
\mathbf{W} &\eqv{\text{DBD},r} f_L(\mathbf{W}),
\nonumber
\\
\mathbf{W} &\eqv{\text{L},r} f'(\mathbf{W}).
\nonumber
\end{align}
\end{linenomath*}
We thus obtain the following results, which completely specify the graphs which are equivalent to the homogeneous matrix $\WH$ for each of our evolutionary dynamics.

\begin{proposition}[Link]
\label{PropositionLink}
$M_{L} = W_C$.
More precisely, the following statements are equivalent:
\begin{enumerate}
\item [(a)] $\mathbf{W}$ is a circulation.
\item[(b)] For all $r>0$, $\mathbf{W} \eqv{L,r} \WH$.
\item[(c)] There is $r>0$ such that $\mathbf{W}\eqv{L,r} \WH.$
\end{enumerate}
\end{proposition}
\noindent

We note that  $W_C = f ' {}^{-1}(W_C) = \{\mathbf{W}; f ' (\mathbf{W})\in W_C\}$ and thus, similarly to Proposition \ref{PropositionBDB} below, Proposition \ref{PropositionLink} can be written as  $M_L = f ' {}^{-1}(W_C)$.

\begin{proposition}[BDB and DBD]
\label{PropositionBDB}
$M_\text{BDB} = f_R^{-1}(W_C)$ and 	$M_\text{DBD} = f_L^{-1}(W_C)$.
More precisely, the following statements are equivalent:
\begin{enumerate}
\item [(a)] $f_R(\mathbf{W})$ is a circulation.
\item[(b)] For all $r>0$, $\mathbf{W} \eqv{\text{BDB},r} \WH$.
\item[(c)] There is $r>0$ such that $\mathbf{W}\eqv{\text{BDB},r} \WH$
\end{enumerate}
The equivalent conditions for DBD are similar to the above for BDB but $f_R$ is replaced by $f_L$.
\end{proposition}

\begin{figure}[h!]
\centering\includegraphics[]{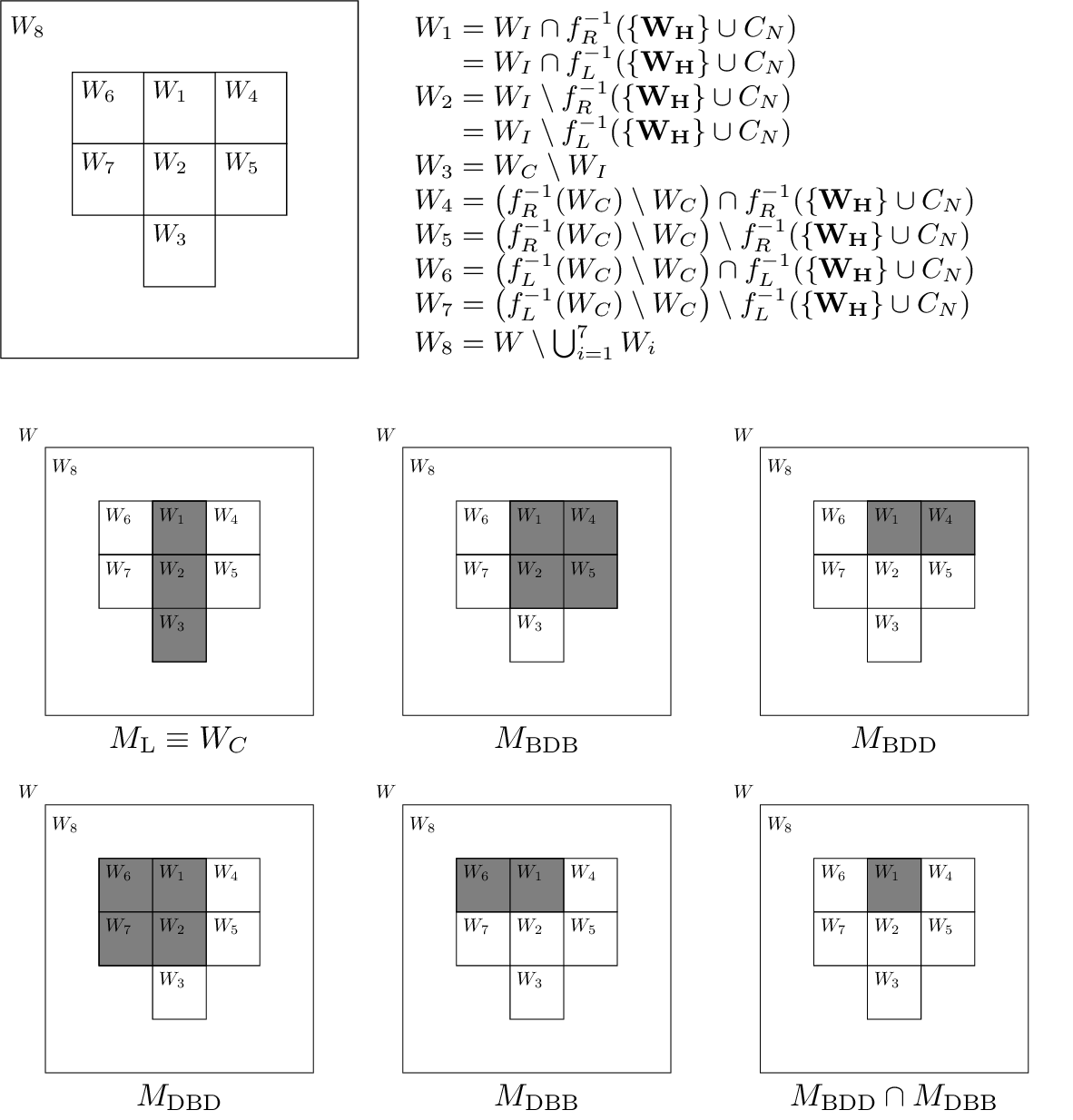}
\caption{
The diagram on top shows eight partitions $W_i$ of $W$ (labelled using their index $i = 1,2,\ldots,8$).
A combination of these partitions make up the sets $M_*$. 
Below this, a separate diagram for each of the standard dynamics is given showing the partitions that make up $M_*$. 
The bottom right diagram shows the partition where $\mathcal{E}$ is $\rho$-equivalent to a Moran process regardless of the standard dynamics on the graph being used, that is, 
$
	M_{\text{L}} \cap 
	M_{\text{BDB}} \cap 	M_{\text{BDD}} \cap 
	M_{\text{DBD}} \cap M_{\text{DBB}}
	\equiv
	M_{\text{BDD}} \cap M_{\text{DBB}} 
$.
\label{fig m sets}
}
\end{figure}

\begin{proposition}[BDD and DBB]
\label{PropositionBDD}
$M_\text{BDB} = f_R^{-1}(\{\mathbf{W_H}\} \cup {C}_N)$ and $M_\text{DBB} = f_L^{-1}(\{\mathbf{W_H}\} \cup {C}_N)$ .
More precisely, the following statements are equivalent:
\begin{enumerate}
\item [(a)] $f_R(\mathbf{W})=\WH$ or $f_R(\mathbf{W})\in C_N$.
\item[(b)] For all $r>0$, $\mathbf{W} \eqv{\text{BDD},r} \WH$.
\end{enumerate}
The equivalent conditions for DBB are similar to the above for BDD but $f_R$ is replaced by $f_L$.
\end{proposition}
\noindent
In particular, $M_\text{BDD}\subset M_\text{BDB}$ and $M_\text{DBB}\subset M_\text{DBD}$.
The sets $M_*$ are illustrated in Figure \ref{fig m sets}.

Note that unlike in Propositions \ref{PropositionLink} and \ref{PropositionBDB},  Proposition \ref{PropositionBDD}
does not contain  ``any $r$ implies all $r$". 
In fact, when $r=1$, there is no selection and thus the dynamics BDB and BDD
are the same (and also the dynamics DBB and DBD are the same). 
Consequently, by Proposition \ref{PropositionBDB},
\begin{linenomath*}
\begin{align}
& 
	\mathbf{W} \eqv{\text{BDD},1} \WH \Leftrightarrow 
	f_R(\mathbf{W})\in W_C \Leftrightarrow 
	\textbf{W} \in M_\text{BDB},
\nonumber
\\
& 
	\mathbf{W} \eqv{\text{DBB},1} \WH \Leftrightarrow 
	f_L(\mathbf{W})\in W_C \Leftrightarrow 
	\textbf{W} \in M_\text{DBD}.
\nonumber
\end{align}
\end{linenomath*}

\subsection{Our results in the context of known results }
For the LB dynamics, Proposition \ref{PropositionLink} was stated and proved in  \cite{2005LiebermanEtal} 
as the  Circulation theorem.
For the LD dynamics, Proposition \ref{PropositionLink} follows from the Circulation theorem and the result of \citep{2009Masuda} that the fixation probabilities for LB and LD are the same.

As shown in Appendix \ref{BDBsameLink},
BDB is the same as the LB dynamics for right stochastic matrices
(in particular, for BDB dynamics, Proposition \ref{PropositionBDB} can be seen as the Isothermal theorem from \cite{2005LiebermanEtal}).
Proposition \ref{PropositionBDB} thus follows from Proposition \ref{PropositionLink} thanks to \eqref{eq:fR equivalence of BDD}. The natural symmetries between $f_R$ and $f_L$ and BDB and DBD dynamics allow us to extend the Isothermal theorem to DBD dynamics as well (see also \citep{2014KavehEtal}).

Overall, Propositions \ref{PropositionLink} and \ref{PropositionBDB} and the occurrence of  $W_C$ within them are consistent with the claim made in \cite{2005LiebermanEtal} that the circulation criterion completely classifies all replacement matrices where $\mathcal{E}_{*,\mathbf{W},r}$ is $\rho$-equivalent to a Moran process.

Our most important new result is Proposition \ref{PropositionBDD}. It shows that the BDD and DBB dynamics require very strict conditions to yield the Moran process. Either the population structure is homogeneous, or it is a directed cycle. This latter structure is an interesting theoretical example, but is unlikely to apply to real populations, meaning that the homogeneous population is practically the only way to get the Moran process for a realistic population. 

\subsection{The importance of self-loops in BDD and DBB dynamics}
Proposition \ref{PropositionBDD}  by definition requires that $w_{ii}>0\ \forall i = 1,2,\ldots,N$. Without such self-loops,
 $\mathcal{E}_{\text{BDD},\mathbf{W},r},\mathcal{E}_{\text{DBB},\mathbf{W},r}$ cannot ever be $\rho$-equivalent to the Moran process.
The ability of an individual to replace itself therefore plays an important role in the replacement structure of the population and cannot be discounted.
For BD dynamics, when increasing the diagonal weights of $\mathbf{W}$, the fixation probability decreases for BDB and increases for BDD.
For DB dynamics, the increase in fixation probability DBB is greater than that for DBD.
For LB dynamics, the fixation probability remains the same.

With BDD and DBD evolutionary dynamics on graphs one may encounter the following problems if there are no self-loops \cite[page 245]{2013_BroomRychtar}.
For DBB dynamics, a type A individual with almost infinite fitness still has a fixation probability bounded away from 1 because even type A individuals can be randomly picked for death and replaced by type B individuals.
With self-loops, however, a type A individual will almost always be replaced by itself (or another type A individual) and therefore has a fixation probability approaching 1.
Similarly, for BDD dynamics, a type A individual with almost zero fitness does not have near probability 0 of fixating as type A  individuals can be randomly picked for birth and replace type B individuals.
With self-loops, such an individual will almost always pick itself (or another type A) to replace and therefore its fixation probability is near 0. 
Thus the inclusion of self-loops removes some problematic features of the  BDD and DBB dynamics, and makes them more attractive dynamics to use in models.

\section{Discussion}

In this paper we have considered an evolutionary graph theory model of a population involving general weights and a variety of evolutionary dynamics based upon the work of \cite{2005LiebermanEtal}, which was a development of the classical population model of \cite{1958Moran}. In such populations, the population size is fixed at all times and at successive discrete time points one replacement event occurs. Like the aforementioned papers we consider two types of individuals, where fitness depends upon type but no other factors (i.e. there are no game-theoretic interactions). In particular the single most important property of such a process is the fixation probability, the probability that a randomly placed mutant individual of the second type will eventually completely replace the population of the first type.

This fixation probability depends upon the fitnesses of the two types of individuals, but can also be heavily influenced by the population structure as given by the weights, and by the evolutionary dynamics used. These effects are commonly observed, although in some circumstances evolution proceeds as if as on a well-mixed population as from the original work of \cite{1958Moran}, dependent only upon the fitnesses of the two types, and some important results in this regard were already given in \cite{2005LiebermanEtal}. The aim of this paper was to provide a generalised set of conditions for when this would be the case.


By defining what is meant by fixation-equivalence to the Moran process, we provided a general result which, independent of the specific dynamics used, helps identify graphs that do not affect the fixation probability.
With respect to each of the standard dynamics, we then classified sets of evolutionary graphs that have the same fixation probability as the Moran process (or well mixed population).
These sets include graphs that are circulations and therefore generalises the work of \cite{2005LiebermanEtal}.

An important new result shows that the set of weights for which we obtain fixation equivalence
to the Moran process for the BDD and DBB dynamics is very restricted, and so that
for most populations with any structure this equivalence will not hold for these dynamics. We
note also that the inclusion of non-zero self weights $w_{ii}$ eliminates some problematic features of
these two dynamics (i.e. that individuals with 0 fitness could fixate or those with infinite fitness
could be eliminated) and so improves the applicability of these dynamics.

Presenting evolutionary dynamics on graphs in the way that we have allows one to
incorporate a variety of dynamics in their analysis, both of standard type and other definitions.
This will improve our understanding of dynamics on graphs in general. 
We note that the list of dynamics in Table \ref{StandardDynamics} is not exhaustive. 
For example, \cite{2006OhtsukiNowak} used imitation
dynamics, which is a class of DBB dynamics with an additional requirement $w_{ii} >0  \ \forall i$.

In general the inclusion of non-zero self weights, in contrast to many earlier evolutionary
graph theory works, allows for a greater flexibility of modelling. We note that this is consistent
with the original work of \cite{1958Moran}, which allowed self-replacement as an integral part of the
process. 
For well-mixed populations it does not matter much whether this possibility is included or not (at least for sufficiently large populations with intermediate fitness values), and it is likely that it has often been excluded for reasons of convenience because of this
without the ramifications being fully considered in many later works. It is thus important to consider whether to include such self weights when modelling spatial structure using evolutionary graph theory.\vskip6pt

\bibliographystyle{vancouver}
\bibliography{Bibliography}

\appendix
\section*{Appendix}
\section{Proofs}


\subsection{BDB is the same as LB for right stochastic matrices}
\label{BDBsameLink}
For BDB dynamics we have $ \mathfrak{r}_{ij}=b_{i}d_{ij}$.
By definition $\sum_{ij}b_id_{ij} =1$, we can therefore write this as $\mathfrak{r}_{ij}={b_{i}d_{ij}}\left/{\sum_{n,k}b_nd_{n,k}}\right.$.
Substituting $b_i={F_i}\left/{\sum_{m=1}^N F_m}\right.$ gives
\begin{linenomath*}\begin{align}
	\mathfrak{r}_{ij}=
	\frac
	{{d_{ij}F_i}\left/{\sum_{m=1}^N F_m}\right.}
	{\sum_{n,k}\left({d_{nk}F_n}\left/{\sum_{m=1}^N F_m}\right.\right)}
	=
	\frac
	{d_{ij}F_i}
	{\sum_{n,k}d_{nk}F_n}
	.
\nonumber
\end{align}\end{linenomath*}
If $\mathbf{W}$ is right stochastic, i.e. $\sum_{n=1}^N w_{in}=1$ for all $i=1,2,\ldots N$, for BDB dynamics we have that $ d_{ij}={w_{ij}}\left/{\sum_{n=1}^N w_{in}}\right.=w_{ij}$ giving 
$\mathfrak{r}_{ij}= {w_{ij}F_i}\left/{\sum_{n,k}w_{nk}F_n}\right.$ 
which is the LB dynamics as required.
We also have that DBD is the same as LD for left stochastic matrices.
The explanation follows the same procedure as above.

\subsection{Lemma \ref{ForwardBias} (Forward Bias)}
The key Lemma \ref{ForwardBias} stated below is used in the proofs of all propositions and it relies heavily on the notion of \emph{forward bias}
 of state $S$ which is then given by the ratio of the probabilities of a forward transition to a backward transition from $S$.
A forward and backward transition from $S$ occurs when the number of type $A$ individuals increase and decrease by one respectively, which happen with probability
\begin{linenomath*}
\begin{align}
	P^+_{S}=
	\sum_{n\notin S} P_{S,S\cup \{n\}} 
	\quad \text{and} \quad
	P^-_S=
	\sum_{n \in S} P_{S,S\setminus\{n\}}.
\nonumber
\end{align}
\end{linenomath*}
\begin{lemma}[Constant Forward Bias]
\label{ForwardBias}
Let $\mathcal{E}$ be an evolutionary process on states $S\subset\{1,2,\ldots, N\}$ with transition probabilities $P_{S,S'}$ that satisfy
\begin{itemize}
 \item  $P_{S,S'}>0$  only if $S$ and $S'$ differ in at most one element
 \item for every $S\neq \emptyset, \{1,\ldots,N\}$, there are $S^+$ and $S^-$ such that $|S^+|=|S|+1$ and $|S^-|=|S|-1$ and $P_{S,S^+}>0, P_{S,S^-}>0$.
 \end{itemize}
 Then, the following are equivalent
\begin{enumerate}
\item[a)]
There is a constant $c>0$ such that for all $S\subset \{1,2,\ldots, N\}$
\begin{linenomath*}
\begin{align*}
	\rho_S^A=
	\begin{cases}
			\displaystyle \frac{1-c^{-|S|}}{1-c^{-N}} & \text{if } c \ne 1,\\
			\displaystyle |S|/N & \text{if } c = 1
	\end{cases}
\end{align*}
\end{linenomath*}
\item[b)]
$\mathcal{E}$ has constant forward bias, that is, there is a constant $d$ such that for all $S\subset \{1,2,\ldots, N\}$
\begin{linenomath*}
\begin{align*}
P^+_S\left/P^-_S\right.=d.
\end{align*}
\end{linenomath*}
\end{enumerate}
Moreover, if either (a) or (b) hold, then $c=d$.
\end{lemma}
Note that a similar result is given in \cite{2005LiebermanEtal,AllenNowak2014} where the forward bias is explicitly defined as
\begin{linenomath*}
\begin{align*} 
r\sum_{a\in S}\sum_{b\notin S} w_{ab}
\left/
	\sum_{a\in S}\sum_{b\notin S} w_{ba}
\right.,
\end{align*}
\end{linenomath*}
which is what one gets when using Link dynamics, or BDB dynamics if $\mathbf{W}\in W_R$.
Note that in Lemma \ref{ForwardBias} the forward bias is defined independent of the dynamics and therefore applies to all dynamics that satisfy the assumptions.

\begin{proof}
	``(a) $\Rightarrow$ (b)":
	Take any $S\subset \{1,2,\ldots, N\}$. It is known that
\begin{linenomath*}
\begin{align*}
	\rho^A_S
		=\sum_{S'}P_{S,S'}\rho^A_{S'}
		=P_{S,S}\rho^A_{S}
		+\sum_{n \notin S}\left(P_{S,S\cup \{n\}}\rho^A_{S\cup \{n\}}\right)
		+\sum_{n \in S}\left(P_{S,S\setminus \{n\}}\rho^A_{S\setminus \{n\}}\right)	
\end{align*}
\end{linenomath*}
and using $P_{S,S}= 1- P^+_S- P^-_S $ gives
\begin{linenomath*}
\begin{align}
	\label{ForwardBiasEq}
	0 =	
		\sum_{n \notin S}\left(P_{S,S\cup \{n\}}\left(\rho^A_{S\cup \{n\}}-\rho^A_{S}\right)\right)
		+\sum_{n \in S}\left(P_{S,S\setminus \{n\}}\left(\rho^A_{S\setminus \{n\}}-\rho^A_{S}\right)\right).
\end{align}
\end{linenomath*}
	For $c\ne 1$, equation \eqref{ForwardBiasEq} simplifies to
\begin{linenomath*}
\begin{align*}
&	
	0 =
	\frac{1-c^{-|S|-1}-1+c^{-|S|}}{1-c^{-N}}	P^+_S
	+
	\frac{1-c^{-|S|+1}-1+c^{-|S|}}{1-c^{-N}} P^-_S
\Rightarrow
\\
&
	{ P^+_S } \left/ { P^-_S }\right.
	=
	\frac{ 	c^{-|S|}-c^{-|S|+1} }
		{ c^{-|S|-1}-c^{-|S|} }			
	=
	\frac{ 1-c } { c^{-1}-1 }
	=
	c.
\end{align*}
\end{linenomath*}
For $c=1$, equation \eqref{ForwardBiasEq} simplifies to
\begin{linenomath*}
\begin{align*}
	0 = ( |S|+1-|S| ) { P^+_S }
	+ ( |S|-1-|S| ) { P^-_S }
	\Rightarrow\quad
	{ P^+_S } \left/ { P^-_S }\right.
	=
	1.
\end{align*}
\end{linenomath*}
	``(b) $\Leftarrow$ (a)'':
	The state transition matrix $\mathbf{S}=(P_{S,S'})$ can be scaled to give $\mathbf{S}'=(P'_{S,S'})$ such that $P'_{S,S}=0$ and $\displaystyle P'_{S,S'}= P_{S,S'} /( 1-P_{S,S} )= P_{S,S'} /( P^+_S + P^-_S )$ where $S$ is a non-absorbing state.
	The fixation probability $\rho^A_S$ will be the same whether $\mathbf{S}'$ or $\mathbf{S}$ is used.
	This is because equation \eqref{Fixation} can be rearranged as follows
\begin{linenomath*}
\begin{align*}
&
	\rho^{A}_S= \sum_{S'} P_{SS'}\rho_{S'}^A
	\Rightarrow\quad 
	\rho^{A}_S=
	P_{SS}\rho_{S}^A +
	\sum_{\substack{S':S'\ne S}}
	P_{SS'}\rho_{S'}^A
	\Rightarrow
\\
&
	\rho^{A}_S (1-P_{SS})=
	\sum_{\substack{S':S'\ne S}}
	P_{SS'}\rho_{S'}^A
	\Rightarrow\quad
	\rho^{A}_S =
	\sum_{\substack{S':S'\ne S}}
	\frac{ P_{SS'} }{  P^+_S + P^-_S }\rho_{S'}^A.
\end{align*}
\end{linenomath*}
	Let $\{\mathcal{S}_0,\mathcal{S}_1,\ldots,\mathcal{S}_N\}$ be a partition of the states $S$ such that $S\in \mathcal{S}_i$ if $|S|=i$.
	The probability $P_{i,j}(S)$ of transitioning from state $S\in \mathcal{S}_i$ to \emph{lumped state} $\mathcal{S}_j$ with respect to $\mathbf{S}'$ is
\begin{linenomath*}
	\begin{align}
		\label{Lumpable}
		P_{i,j}(S)=
		\begin{cases}
			0
		&
			j \ne i\pm1,
		\\
			{1}/(d+1)
		&
			j= i-1,
		\\
			{d}/(d+1)
		&
			j=i+1
		\end{cases}
		\quad \text{for } i=1,2,\ldots,N-1.
	\end{align}
\end{linenomath*}
	This can be easily verified, for example, take $j=i-1$ then
\begin{linenomath*}
	\begin{align*}
	&
		P_{i,i-1}(S)
	=
		\sum_{S'\in \mathcal{S}_{i-1}}P'_{S,S'}
	=
		\sum_{S'\in \mathcal{S}_{i-1}} \frac{P_{S,S'}}{P^+_S + P^-_S}
	=
		\frac{P^-_S}{P^+_S + P^-_S}
	=
		\frac{1}{1+d}
	\end{align*}
\end{linenomath*}
	since the forward bias is equal to $d$.
	Equation \eqref{Lumpable} satisfies the necessary and sufficient condition for the Markov chain with state transition matrix $\mathbf{S}'$ to be lumpable with respect to the partition $\{\mathcal{S}_0,\mathcal{S}_1,\ldots,\mathcal{S}_N\}$  (Theorem 6.3.2 page 124, \cite{1960KemenySnell}).
	Let $\hat{\mathbf{S}}=(P_{i,j})$ be the state transition matrix for this lumped Markov chain then the probability $P_{i,j}$ of transitioning from lumped states $\mathcal{S}_i$ to $\mathcal{S}_j$ is given by
\begin{linenomath*}
	\begin{align*}
		P_{i,j}=P_{i,j}(S).
	\end{align*}
\end{linenomath*}
	The state transition matrix $\hat{\mathbf{S}}$ describes a random walk with absorbing barriers and therefore the probability $\rho^A_{i}$ of type $A$ individuals fixating when the population starts in lumped state $\mathcal{S}_i$ is calculated using the methods in \cite{1975KarlinTaylor} to give
\begin{linenomath*}
	\begin{align*}
		\rho^A_{i}=
			{
			\displaystyle 1+\sum_{j=1}^{i-1}\prod_{k=1}^{j}
			\frac{P_{k,k-1}}
				{P_{k,k+1}}
			}\left/{
			\displaystyle 1+\sum_{j=1}^{N-1}\prod_{k=1}^{j}
			\frac{P_{k,k-1}}
				{P_{k,k+1}}
			}\right..
	\end{align*}
\end{linenomath*}
	In this case,
\begin{linenomath*}
	\begin{align*}
		\rho^A_{i}=
		\begin{cases}
			\displaystyle
			\frac{1-d^{-i}}{1-d^{-N}}
		&
			d \ne 1,	
		\\
			i/N
		&
			d = 1
		\end{cases}
	\end{align*}
\end{linenomath*}
	since $P_{k,k-1}/P_{k,k+1}=1/r$ for $k=1,2,\ldots,N-1$.
	By definition, $\rho^A_S=\rho^A_{i}$ where $i=|S|$ as required.
\end{proof}

\subsection{Proposition \ref{PropositionLink} (Link)}
The following statements are equivalent:
\begin{enumerate}
\item [(a)] $\mathbf{W}$ is a circulation.
\item[(b)] For all $r>0$,
$\mathbf{W} \eqv{\text{L},r} \WH$.
\item[(c)] There is $r>0$ such that
$\mathbf{W}\eqv{\text{L},r} \WH.$
\item [(d)] For all $r>0$ and for all $S\subset\{1,2,\ldots, N\}$, the forward bias of $\mathcal{E}_{\text{L},\mathbf{W},r}$ is $r$, i.e.
\begin{linenomath*}
\begin{align}
{ P^+_S } \left/ { P^-_S }\right. = r.
\nonumber
\end{align}
\end{linenomath*}
\item [(e)] There is $r>0$ such that for all $a\in \{1,2,\ldots, N\}$, the forward bias of the one element set $S=\{a\}$ is $r$, i.e.
\begin{linenomath*}
\begin{align}\frac{	
	\displaystyle	\sum_{b \neq a } P_{\{a\},\{a,b\}}
}{
	\displaystyle	 P_{a,\emptyset}	
} = r.
\nonumber
\end{align}
\end{linenomath*}

\end{enumerate}

\begin{proof}
	For LB dynamics the forward bias is given by
\begin{linenomath*}
\begin{align*}
\frac{ P^+_S }{ P^-_S }
=
\frac{
	\displaystyle
	\sum_{a\in S}
	\sum_{b \notin S}
	\frac{
		\displaystyle	
		w_{ab}F_a
	}
	{
		\displaystyle \sum_{n,k}w_{nk}F_n
	}
}
{
	\displaystyle	
	\sum_{a\in S}
	\sum_{b \notin S}
	\frac{
		\displaystyle	
		w_{ba}F_b
	}
	{
		\displaystyle
		\sum_{n,k}w_{nk}F_n
	}
}
=
\frac{
	\displaystyle
	r
	\sum_{a\in S}
	\sum_{b \notin S}
	w_{ab}
}
{
	\displaystyle	
	\sum_{a\in S}
	\sum_{b \notin S}
	w_{ba}
}.
\end{align*}
\end{linenomath*}
For LD dynamics the forward bias is given by
\begin{linenomath*}
\begin{align*}
\frac{ P^+_S }{ P^-_S }
=
\frac{
	\displaystyle
	\sum_{a\in S}
	\sum_{b \notin S}
	\frac{
		\displaystyle	
		w_{ab}/F_b
	}
	{
		\displaystyle \sum_{n,k}w_{nk}/F_k
	}
}
{
	\displaystyle	
	\sum_{a\in S}
	\sum_{b \notin S}
	\frac{
		\displaystyle	
		w_{ba}/F_a
	}
	{
		\displaystyle
		\sum_{n,k}w_{nk}/F_k
	}
}
=
\frac{
	\displaystyle
	r
	\sum_{a\in S}
	\sum_{b \notin S}
	w_{ab}
}
{
	\displaystyle	
	\sum_{a\in S}
	\sum_{b \notin S}
	w_{ba}
}.
\end{align*}
\end{linenomath*}
\noindent
	``(a) $\Rightarrow$ (d)'': $\mathbf{W}$ is a circulation
i.e. $T^+_n=T^-_n$ \ for all $n\in \{1,\ldots, N\}$ and thus
\begin{linenomath*}
\begin{align*}
	\sum_{a\in S}	\sum_{b \notin S}	w_{ab}
&=
	\sum_{a\in S}
	\bigg(
		\sum_n w_{an} -	\sum_{k\in S} w_{ak}
	\bigg)
=		
    \sum_{a\in S}
	\bigg( 
		T^+_a -	 \sum_{k\in S} w_{ak}
	\bigg) 
	\Rightarrow
\\
	\sum_{a\in S}	\sum_{b \notin S}	w_{ab}
&=		
	\sum_{a\in S} 
	\bigg(
		T^-_a - 	\sum_{k\in S}	w_{ka} 
	\bigg)
	= 		
	\sum_{a\in S}
	\bigg(
		\sum_n w_{na} -
		\sum_{k\in S} w_{ka}
	\bigg)
	\Rightarrow
\\
	\sum_{a\in S}	\sum_{b \notin S}	w_{ab}
&=
   \sum_{a\in S} \sum_{b \notin S}	w_{ba}.
\end{align*}
\end{linenomath*}
Note that $ \sum_{a\in S} 			
\sum_{b \notin S} w_{ab}\neq 0$ because $\mathbf{W}$ is admissible and represents a strongly connected graph.
Thus,  the forward bias for both LB and LD is equal to $r$.
\\
``(d)$\Rightarrow$(e)" is trivial as (d) is much stronger than
(e).
\\
``(e)$\Rightarrow$(a)" Let $a$ and $r$ is fixed. By above
calculations of the forward bias, we have
\begin{linenomath*}
\begin{align*}
	\sum_{b \notin S=\{a\}}
	w_{ab}
	=
	\sum_{b \notin S=\{a\}}
	w_{ba}
\Rightarrow\quad
	-w_{aa}
	+\sum_{i = 1}^N
	w_{ai}
	=
	-w_{aa}
	+\sum_{i = 1}^N
	w_{ia}
\Rightarrow\quad
	\sum_{i = 1}^N
	w_{ai}
	=
	\sum_{i = 1}^N
	w_{ia}
\end{align*}
\end{linenomath*}
therefore $\mathbf{W}$ is a circulation.
\\
``(d)$\Rightarrow$(b)" follows from Lemma \ref{ForwardBias}.\\
``(b)$\Rightarrow$(c)" is trivial.\\
``(c)$\Rightarrow$(e)" follows from
 Lemma \ref{ForwardBias}.
\qedhere
\end{proof}

\subsection{Proposition \ref{PropositionBDB} (BDB and DBD)}
More precisely, the following statements are equivalent:
\begin{enumerate}
\item [(a)] $f_R(\mathbf{W})$ is a circulation.
\item[(b)] For all $r>0$,
$\mathbf{W} \eqv{\text{BDB},r} \WH$.
\item[(c)] There is $r>0$ such that
$\mathbf{W}\eqv{\text{BDB},r} \WH$
\item [(d)] For all $r>0$ and for all $S\subset\{1,2,\ldots, N\}$, the forward bias of $\mathcal{E}_{\text{BDB},\mathbf{W},r}$ is $r$, i.e.
\begin{linenomath*}
\begin{align}
{ P^+_S } \left/ { P^-_S }\right. = r.
\nonumber
\end{align}
\end{linenomath*}
\item [(e)] There is $r>0$ such that for all $a\in \{1,2,\ldots, N\}$, the forward bias of $\mathcal{E}_{\text{BDB},\mathbf{W},r}$  of the one element set $S=\{a\}$ is $r$, i.e.
\begin{linenomath*}
\begin{align}\frac{	
	\displaystyle	\sum_{b \neq a } P_{\{a\},\{a,b\}}
}{
	\displaystyle	 P_{a,\emptyset}	
} = r.
\nonumber
\end{align}
\end{linenomath*}
\end{enumerate}
\noindent

\begin{proof}
	Let $\mathbf{U}=(u_{ij})=f_R(\mathbf{W})=\left(w_{ij}/\sum_{n}w_{in}\right)$ then for BDB dynamics the forward bias of $\mathcal{E}_{BDB, \mathbf{W},r}$ is given by
\begin{linenomath*}
\begin{align}
\frac{ P^+_S }{ P^-_S }
=
\frac
	{
		\displaystyle
	\sum_{a\in S}
	\sum_{b \notin S}
		\frac{F_{a}}{\displaystyle \sum_n F_n}
		\frac{\displaystyle w_{ab}}{\displaystyle \sum_n w_{an}}				
	}
	{
		\displaystyle
	\sum_{a\in S}
	\sum_{b \notin S}
		\frac{F_{b}}{\displaystyle \sum_n F_n}
		\frac{\displaystyle w_{ba}}{\displaystyle \sum_n w_{bn}}
	}
=
\frac
	{
		\displaystyle
		r
		\sum_{a\in S}
		\sum_{b \notin S}
		u_{ab}
	}
	{
		\displaystyle
		\sum_{b\notin S}
		\sum_{a \in S}
		u_{ba}
	}
\nonumber
\end{align}
\end{linenomath*}
and therefore the forward bias of $\mathcal{E}_{\text{BDB},\mathbf{W},r}$ is the same as forward bias of $\mathcal{E}_{\text{BDB},\mathbf{U},r}$.

Similarly, with almost identical working as above, when $\mathbf{V} = f_L(\mathbf{W})$, the forward bias of 
$\mathcal{E}_{\text{DBD},\mathbf{W},r}$ is the same as forward bias of $\mathcal{E}_{\text{DBD},\mathbf{V},r}$
and is given by

\begin{linenomath*}
\begin{align*}
\frac{ P^+_S }{ P^-_S }
		=\frac{
			\displaystyle
			\sum_{a\in S}	\sum_{b \notin S}
			\frac{1/F_b}{\displaystyle\sum_n 1/F_n}
			\frac{w_{ab}}
			{
				\displaystyle
				\sum_n w_{nb}
			}
		}
		{
			\displaystyle
			\sum_{a\in S}	\sum_{b \notin S}
			\frac{1/F_a}{\displaystyle\sum_n 1/F_n}
			\frac{w_{ba}}
			{
				\displaystyle
				\sum_n w_{na}
			}
		}
		=
		\frac{
			\displaystyle
			\sum_{a\in S} \sum_{b\notin S}v_{ab}
		}
		{
			\displaystyle
			\frac{1}{r} \sum_{a\in S} \sum_{b\notin S} v_{ba}
		}.
	\end{align*}
\end{linenomath*}
and the proof of the Proposition for DBD closely follows the one for BDB given below with $\mathbf{U}$ and $f_R$ appropriately replaced by $\mathbf{V}$ and $f_L$.

	``(a)$\Rightarrow$ (d)'': If $\mathbf{U}=f_R(\mathbf{W})\in W_\text{C}$, i.e. if $\mathbf{U}$ is doubly stochastic, then the forward bias (for $S\neq \emptyset, \mathcal{N}$) is equal to
\begin{linenomath*}
	\begin{align*}
		 \frac{P^+_S}{P^-_S}=
		\frac
			{
				\displaystyle
				r
				\sum_{a\in S}
				\bigg(\sum_{n}(u_{an}) -\sum_{k \in S} (u_{ak})\bigg)
			}
			{
				\displaystyle
				\sum_{a\in S}
				\bigg(\sum_{n}(u_{na}) -\sum_{k \in S} (u_{ka})\bigg)
			}
		=
		\frac
			{
				\displaystyle
				r
				\bigg(|S| -\sum_{a\in S}\sum_{k \in S} u_{ak}\bigg)
			}
			{
				\displaystyle
				|S| -\sum_{a\in S}\sum_{k \in S} u_{ka}
			}
		=r		
	\end{align*}
\end{linenomath*}
\\
``(d)$\Rightarrow$(e)" is trivial as (d) is stronger than
(e).
\\
``(e)$\Rightarrow$(a)" Let $a$ and $r$ is fixed. By above
calculations of the forward bias, we have
\begin{linenomath*}
	\begin{align*}
		\displaystyle
		\sum_{a\in S}
		\sum_{b \notin S}
		u_{ab}
		=&
		\sum_{a\in S}
		\sum_{b \notin S}
		u_{ba}.
	\end{align*}
\end{linenomath*}
	Consider the states $S=\{a\}$ in which there is only one individual of type $A$ then
\begin{linenomath*}
	\begin{align*}
		\sum_{b \notin S} u_{ab}				
		=
		\sum_{b \notin S} u_{ba}
		\Rightarrow\quad
		- u_{aa}	+\sum_{i = 1}^N u_{ai}
		=
		-u_{aa} +\sum_{i = 1}^N u_{ia}
		\Rightarrow\quad
		1=\sum_{i = 1}^N u_{ia}
	\end{align*}
\end{linenomath*}
	is true for all $a=1,2,\ldots,N$ and therefore $\mathbf{U}$ is doubly stochastic  and thus $f_R(\mathbf{W})$ is a circulation.
\\
``(d)$\Rightarrow$(b)" follows from Lemma \ref{ForwardBias}.\\
``(b)$\Rightarrow$(c)" is trivial.\\
``(c)$\Rightarrow$(e)" follows from
 Lemma \ref{ForwardBias}.
\qedhere
\end{proof}

\subsection{Proposition \ref{PropositionBDD} (BDD and DBB)}
The following statements are equivalent:
\begin{enumerate}
\item [(a)] $f_R(\mathbf{W})=\WH$ or $f_R(\mathbf{W})\in C_N$.
\item[(b)] For all $r>0$, $\mathbf{W} \eqv{\text{BDD},r} \WH$.
\end{enumerate}
\begin{proof}
The replacement probabilities $\mathfrak{r}_{ij}(\mathbf{F}(S),\mathbf{W})$ for BDD dynamics can be rewritten as $\mathfrak{r}_{ij}(\mathbf{F}(S),\mathbf{U})$ where $\mathbf{U}=(u_{ij})=f_R(\mathbf{W})=\left(w_{ij}/\sum_{n}w_{in}\right)$ by multiplying the numerator and denominator with $\sum_{n}w_{in}$ as follows

\begin{linenomath*}\begin{align}
&
	\mathfrak{r}_{ij}(\mathbf{F}(S),\mathbf{W}) =
	\frac{1}{N}\frac{w_{ij}/F_{j}(S)}{ \sum_n w_{in}/F_n(S)}=
	\frac{1}{N}\frac{w_{ij}/\left(F_j(S)\sum_n w_{in}\right)}{\sum_n w_{in}/\left(F_n(S)\sum_n w_{in}\right)}\Rightarrow
\nonumber
\\
&
	\frac{u_{ij}/F_j(S)}{ \sum_n u_{in}/F_n(S)}
	=\mathfrak{r}_{ij}(\mathbf{F}(S),\mathbf{U})
\nonumber
\end{align}\end{linenomath*}
and therefore we have that $\mathbf{W}\eqv{\text{BDD},r}\mathbf{U}$, for all $r>0$.
The forward bias using $\mathbf{U}$ for state $S$ is given by
\begin{linenomath*}\begin{align}
\frac{ P^+_S }{ P^-_S }
=
\frac
	{
		\displaystyle
		\sum_{a\in S}	\sum_{b \notin S}
		\frac{1}{N}
		\frac{\displaystyle u_{ab}/F_b}{\displaystyle \sum_n u_{an}/F_n}				
	}
	{
		\displaystyle
		\sum_{a\in S}	\sum_{b \notin S}
		\frac{1}{N}
		\frac{\displaystyle u_{ba}/F_a}{\displaystyle \sum_n u_{bn}/F_n}
	}
=
\frac
	{
		\displaystyle
			\sum_{a\in S}	\sum_{b \notin S}
		\frac{\displaystyle u_{ab}}{\displaystyle \sum_n u_{an}/F_n}				
	}
	{
		\displaystyle
		\frac{1}{r}
			\sum_{a\in S}	\sum_{b \notin S}
		\frac{\displaystyle u_{ba}}{\displaystyle \sum_n u_{bn}/F_n}
	}.
\label{ProofBDDFB}
\end{align}\end{linenomath*}

Similarly, let $\mathbf{V}=(v_{ij})=f_L(\mathbf{W})=(w_{ij}/\sum_n w_{nj})$. Then for DBB dynamics we have
\begin{linenomath*}
\begin{align}
	b_{ij} = \frac{w_{ij}F_i}{ \sum_n w_{nj}F_n}
	= \frac{w_{ij}F_i/\sum_n w_{nj}}{\sum_n w_{nj}F_n/\sum_n w_{nj} }
	= \frac{v_{ij}F_i}{\sum_n v_{nj}F_n }
\nonumber
\end{align}
\end{linenomath*}
and therefore the forward bias when using $\mathbf{V}$ is given by
\begin{linenomath*}
\begin{align}
\frac{ P^+_S }{ P^-_S }
	=\frac{
		\displaystyle
			\sum_{a\in S}	\sum_{b \notin S}
		\frac{1}{N}
		\frac{v_{ab}F_a}
		{
			\displaystyle
			\sum_n v_{nb}F_n
		}
	}
	{
		\displaystyle
			\sum_{a\in S}	\sum_{b \notin S}
		\frac{1}{N}
		\frac{v_{ba}F_b}
		{
			\displaystyle
			\sum_n v_{na}F_n
		}
	}
	=
	\frac{
		\displaystyle
		r
			\sum_{a\in S}	\sum_{b \notin S}
		\frac{v_{ab}}
		{
			\displaystyle
			\sum_n v_{nb}F_n
		}
	}
	{
		\displaystyle
			\sum_{a\in S}	\sum_{b \notin S}
		\frac{v_{ba}}
		{
			\displaystyle
			\sum_n v_{na}F_n
		}
	}.
\nonumber
\end{align}
\end{linenomath*}
The proof of the Proposition  for DBB closely follows the one for BDD given below with $\mathbf{U}$ and $f_R$ appropriately replaced by $\mathbf{V}$ and $f_L$.

\subsubsection{If $\mathbf{U}\in C_N$, then $\mathbf{U} \eqv{\text{BDD},r} \WH$}
If $\mathbf{U}\in C_N$ then there are only two nonzero elements in each row.
In particular, in row $i$ of $\mathbf{U}$ we have that $u_{ii},u_{ik_i}=1/2$ for some $k_i \ne i$.
In the numerator of equation \eqref{ProofBDDFB} for $a\in S$, $b \notin S$ and $k_a \ne a$ we have that for all $S$
\begin{linenomath*}\begin{align}
&
	\frac{u_{ab}}{\displaystyle\sum_n u_{an}/F_n(S)}
	=
	\frac{u_{ab}}{ u_{aa}/F_a(S) + u_{ak_a}/F_{k_a}(S)}
	=
	\begin{cases}
		0 & \text{if } b \ne k_a, \\
		\frac{1/2}{1/2r + 1/2} & \text{if } b = k_a.
	\end{cases}
\nonumber
\end{align}\end{linenomath*}
Similarly, in the denominator of equation \eqref{ProofBDDFB} for $a\in S$, $b \notin S$ and $k_b \ne b$ we have that for all $S$
\begin{linenomath*}\begin{align}
&
	\frac{u_{ba}}{\displaystyle\sum_n u_{bn}/F_n(S)}
	=
	\frac{u_{ba}}{ u_{bb}/F_b(S) + u_{bk_b}/F_{k_b}(S)}
	=
	\begin{cases}
		0 & \text{if } a \ne k_b, \\
		\frac{1/2}{1/2 + 1/2r} & \text{if } a = k_b.
	\end{cases}
\nonumber
\end{align}\end{linenomath*}
This means that equation \eqref{ProofBDDFB} for all $S$ can be written as
\begin{linenomath*}\begin{align}
	\frac{x/2}{1/2r+1/2}\left/\frac{1}{r}\frac{y/2}{1/2+1/2r}\right.=rx/y
\nonumber
\end{align}\end{linenomath*} where $x$ ($y$) is the number of nonzero $u_{ab}$ ($u_{ba}$) terms in the numerator (denominator).
If we partition the vertices of the digraph of $\mathbf{U}$ into any two sets $V_1,V_2$ then the number of edges $e(i,j)$ and $e(j,i)$ for $i \in V_1$ and $j\in V_2$ are by definition the same because it is a cycle.
This means that for $a \in S$ and $b\notin S$ the number of nonzero $u_{ab},u_{ba}$ terms in the numerator and denominator respectively are the same hence $x=y$ and $rx/y=r$ as required.
As per 	Lemma \ref{ForwardBias}, $\mathcal{E}_{\text{BDD},\mathbf{U},r}$ is $\rho$-equivalent to the Moran process.

\noindent
\subsubsection{If $\mathbf{U} \eqv{\text{BDD},r} \WH$ for all $r>0$, then $\mathbf{U}=\WH$ or $\mathbf{U} \in C_N$}
By  Lemma \ref{ForwardBias}, the forward bias \eqref{ProofBDDFB} is equal to $r$ for all $S\subset\{1, \ldots, N\}$ giving
\begin{linenomath*}\begin{align}
&
	\sum_{a\in S}	\sum_{b \notin S}
	\frac{\displaystyle u_{ab}}{\displaystyle \sum_n u_{an}/F_n}				
	=
	\sum_{a\in S}	\sum_{b \notin S}
	\frac{\displaystyle u_{ba}}{\displaystyle \sum_n u_{bn}/F_n}
\Rightarrow
\nonumber
\\
&
	\sum_{a\in S}	
	\frac
		{
			\displaystyle
			\sum_{b \notin S} u_{ab}
		}
		{
			\displaystyle
			\sum_{j \notin S} u_{aj}+
			\frac{1}{r} \sum_{i \in S} u_{ai}
		}				
	=
	\sum_{b \notin S}
	\frac
		{
			\displaystyle
			\sum_{a \in S} u_{ba}
		}
		{
			\displaystyle
			\sum_{j \notin S} u_{bj}+
			\frac{1}{r} \sum_{i \in S} u_{bi}
		}
\label{ProofBDDForwardBiasr}
.
\end{align}\end{linenomath*}
Note that if $r=1$, \eqref{ProofBDDForwardBiasr} holds for all $\mathbf{U}\in W_C$. 
From now, we will consider $r\neq 1$ only.
For clarity, the remainder of this section of the proof is broken down into the following six steps.

\newcounter{steps}
\stepcounter{steps}
\subsubsection*{Step \arabic{steps}: Derivation of general state dependent row-sum equation}
Let $U(a,S)=\sum_{i \in S} u_{ai}$, i.e. $1-U(a,S)=\sum_{j \notin S } u_{aj}$.
Equation \eqref{ProofBDDForwardBiasr} thus becomes
\begin{linenomath*}\begin{align}
&
	\sum_{a\in S}	
	\frac
		{
			\displaystyle
			1 - U(a,S)
		}
		{
			\displaystyle
			1 - U(a,S) + U(a,S)/r
		}				
	=
	\sum_{b \notin S}
	\frac
		{
			\displaystyle
			U(b,S)
		}
		{
			\displaystyle
			1-U(b,S)+U(b,S)/r
		}
\Rightarrow
\nonumber
\\
&
	\sum_{a\in S}
	\frac
		{
			\displaystyle
			1
		}
		{
			\displaystyle
			1 + U(a,S)(1/r-1)
		}				
	=
	\sum_{n=1}^N
	\frac
		{
			\displaystyle
			U(n,S)
		}
		{
			\displaystyle
			1+U(n,S)(1/r-1)
		}
\label{ProofBDDForwardBiasrSimplified}
.
\end{align}\end{linenomath*}
Equation \eqref{ProofBDDForwardBiasrSimplified} can be written as a Taylor series as follows
\begin{linenomath*}\begin{align}
&
	\sum_{a\in S}\sum_{k=0}^{\infty} (-1)^k(1/r-1)^k\left[U(a,S)\right]^k
	=
	\sum_{n=1}^N U(n,S) \sum_{k=0}^{\infty} (-1)^k(1/r-1)^k\left[U(n,S)\right]^k
\Rightarrow
\nonumber
\\
&
	\sum_{a\in S}\sum_{k=0}^{\infty} (1-1/r)^k\left[U(a,S)\right]^k
	=
	\sum_{n=1}^N \sum_{k=0}^{\infty} (1-1/r)^k\left[U(n,S)\right]^{k+1}
\label{ProofBDDTaylorSeries}
\end{align}\end{linenomath*}
For equation \eqref{ProofBDDTaylorSeries} to hold for all $r$ the coefficients of $(1-1/r)^k$ should be same, that is, for all $k$
\begin{linenomath*}\begin{align}
&
	\sum_{a\in S} \left[U(a,S)\right]^k
	=
	\sum_{n = 1}^N \left[U(n,S)\right]^{k+1}
.
\label{ProofBDDTaylorSeriesCoefficients}
\end{align}\end{linenomath*}

\stepcounter{steps}
\subsubsection*{Step \arabic{steps}: The diagonal of $\mathbf{U}$ consists of non-zero elements}
Consider the state $S=\{a\}$ then equation \eqref{ProofBDDTaylorSeriesCoefficients} gives
\begin{linenomath*}\begin{align}
&
	u_{aa}^k
	=
	\sum_{n = 1}^N u_{na}^{k+1}.
\label{ProofBDDTaylorOneA}
\end{align}\end{linenomath*}	
If $u_{aa}=0$ or 1 then \eqref{ProofBDDTaylorOneA} implies that all off-diagonal terms in column $n$ are zero which is a contradiction with  $\mathbf{W}$ (and thus also $\mathbf{U}=f_R(\mathbf{W})$) being strongly connected, which means that $0<u_{aa}<1$.

\stepcounter{steps}
\subsubsection*{Step \arabic{steps}: The $n^\text{th}$ column of $\mathbf{U}$ contains $m_n$ nonzero elements, all equal to $1/m_n$}
Since $0<u_{aa}<1$, we can divide equation \eqref{ProofBDDTaylorOneA} by $u_{aa}^k$ giving
\begin{linenomath*}\begin{align}
	1 =
	\sum_{ n = 1 }^N u_{ na }
	\left(\frac{u_{na}}{u_{aa}}\right)^{k}
\label{ProofBDDTaylorOneASimp}
.
\end{align}\end{linenomath*}
We have that
\begin{linenomath*}\begin{align}
	\lim_{k\to\infty}
	\left(\frac{u_{na}}{u_{aa}}\right)^{k}
	=
	\begin{cases}
		\infty 	& u_{na}>u_{aa},\\
		1 		& u_{na}=u_{aa},\\
		0 		& u_{na}<u_{aa},
	\end{cases}
\nonumber
\end{align}\end{linenomath*}
and therefore \eqref{ProofBDDTaylorOneASimp} implies that $0\le u_{na}\le u_{aa}$.
There must be $n\neq a$ such that  $u_{na}= u_{aa}$ as otherwise, by \eqref{ProofBDDTaylorOneASimp}, we  would have  $u_{aa}=1$.
Let $\mathcal{C}_a=\{i:u_{ia}=u_{aa}\}$. 
\eqref{ProofBDDTaylorOneASimp} becomes
\begin{linenomath*}\begin{align}
	1&=
		\bigg(\sum_{i\in C_a} u_{aa}\bigg)+
		\bigg(\sum_{j\notin C_a} \frac{u_{ja}^{k+1}}{u_{aa}^k}\bigg)
	=
		|C_a|u_{aa}+
		\bigg(\sum_{j\notin C_a} \frac{u_{ja}^{k+1}}{u_{aa}^k}\bigg).
\label{eq: for M_a}
\end{align}\end{linenomath*}
As $k\to\infty$, \eqref{eq: for M_a} implies that $u_{aa}=1/|\mathcal{C}_a|$.
Thus, again by \eqref{eq: for M_a}, $u_{ja}=0$ for all $j\notin \mathcal{C}_a$.
This means that in column $n$ of $\mathbf{U}$ there should be $m_n = |C_n|$ with $2\leq m_n\leq N$ nonzero elements, including $u_{nn}$, that are all equal to $1/m_n$.

\stepcounter{steps}
\subsubsection*{Step \arabic{steps}: $m_n$ is the same for all $n$}
Considering state $S=\{i,j\}$ and using $u_{aa}=1/m_a$,
\eqref{ProofBDDTaylorSeriesCoefficients} can be written as  follows
\begin{linenomath*}\begin{align}
	(u_{ii} + u_{ij})^k+ (u_{ji} + u_{jj})^k
	=
&
	\alpha \frac{1}{m_i^{k+1}}+
	\beta \frac{1}{m_j^{k+1}}+
	\gamma \left(\frac{1}{m_{i}}+\frac{1}{m_{j}}\right)^{k+1}
\label{ProofBDDTaylorTwoA}
\end{align}\end{linenomath*}
where $\alpha, \beta, \gamma$ are the number of rows where $1/m_i$ is adjacent to 0, 0 is adjacent to $1/m_j$, and $1/m_i$ is adjacent to $1/m_j$ in columns $i$ and $j$ respectively.
More precisely,
	$\alpha$ is the cardinality of the set $K_{ij} ^{i}= \{n: u_{ni}=1/m_i, u_{nj}=0\}$,
	$\beta$ is the cardinality of the set $K_{ij}^j = \{n: u_{ni}=0, u_{nj}=1/m_j\}$ and
	$\gamma$ is the cardinality of the set $K_{ij}^{ij} = \{n: u_{ni}=1/m_i, u_{nj}=1/m_j\}$.

Since $\mathcal{C}_i= K_{ij}^i\cup K_{ij}^{ij}$ and $\mathcal{C}_j=K_{ij}^j\cup K_{ij}^{ij}$,  we have that $m_i=\alpha+\gamma$ and $m_j=\beta+\gamma$.
Since $K_{ij}^{i},K_{ij}^{j},K_{ij}^{ij}$ are disjoint, we have $\alpha+\beta+\gamma \le N$.
Now, consider the different possibilities we can have on the left-hand side of equation \eqref{ProofBDDTaylorTwoA}.
\vspace{0.2cm}

\noindent
\textbf{\emph{Case 1:}}\\
 $u_{ii}=1/m_i,\ u_{ij}=0$ in row $i$ and $u_{ji}=1/m_i,\ u_{jj}=1/m_j$ in row $j$.
 Thus $\alpha,\gamma \ge 1$ and therefore equation \eqref{ProofBDDTaylorTwoA} gives
\begin{linenomath*}\begin{flalign}
&
	\frac{1}{m_i^k} + \left(\frac{m_i+m_j}{m_i m_j}\right)^k
	=
	\frac{\alpha}{m_i^{k+1}}+
	\frac{\beta}{m_j^{k+1}}+
	\gamma \left(\frac{m_i+m_j}{m_i m_j}\right)^{k+1}
\Rightarrow
\nonumber
\\
&
	\frac{1}{(\alpha+\gamma)^k} + \left(\frac{\alpha+\beta+2\gamma}{(\alpha+\gamma)(\beta+\gamma)}\right)^k
	=
	\frac{\alpha}{(\alpha+\gamma)^{k+1}}+
	\frac{\beta}{(\beta+\gamma)^{k+1}}+
	\gamma\left(\frac{\alpha+\beta+2\gamma}{(\alpha+\gamma)(\beta+\gamma)}\right)^{k+1}
\Rightarrow
\nonumber
\\
&
	\frac
	{
		(\beta+\gamma)^{k} + (\alpha+\beta+2\gamma)^{k}
	}
	{
		[(\alpha+\gamma)(\beta+\gamma)]^{k}
	}
	=
	\frac
	{
		\alpha(\beta+\gamma)^{k+1} + \beta(\alpha+\gamma)^{k+1} + \gamma(\alpha+\beta+2\gamma)^{k+1}
	}
	{
		[(\alpha+\gamma)(\beta+\gamma)]^{k+1}
	}
\Rightarrow
\nonumber
\\
&
	(\beta+\gamma)^{k} + (\alpha+\beta+2\gamma)^{k}
	=
	\frac
	{
		\alpha(\beta+\gamma)^{k+1} + \beta(\alpha+\gamma)^{k+1} + \gamma(\alpha+\beta+2\gamma)^{k+1}
	}
	{
		(\alpha+\gamma)(\beta+\gamma)
	}
\Rightarrow
\nonumber
\\
&
	(\beta+\gamma)^{k} + (\alpha+\beta+2\gamma)^{k}
	=
	\frac{\alpha(\beta+\gamma)^{k}}{\alpha+\gamma}+
	\frac{\beta(\alpha+\gamma)^{k}}{\beta+\gamma}+
	\frac{(\alpha\gamma+\beta\gamma+2\gamma^2)(\alpha+\beta+2\gamma)^{k}}{\alpha\beta+\alpha\gamma+\beta\gamma +\gamma^2}
\Rightarrow
\nonumber
\\
&
	\frac{\gamma(\beta+\gamma)^{k} }{\alpha+\gamma}
	=
	\frac{\beta(\alpha+\gamma)^{k}}{\beta+\gamma}+
	\frac{(\gamma^2-\alpha\beta)(\alpha+\beta+2\gamma)^{k}}{\alpha\beta+\alpha\gamma+\beta\gamma+\gamma^2}.
\nonumber
&
\end{flalign}\end{linenomath*}
As $k\to \infty$, we get
\mbox{$(\beta + \gamma )^{k} \ne	(\alpha+\gamma )^{k} \pm (\alpha+\beta +2\gamma )^{k}$}
since $\alpha + \beta + 2\gamma >\beta + \gamma ,\ \alpha + \gamma $ hence we want $\gamma^2=\alpha\beta$ to get rid off $(\alpha+\beta+2\gamma )^{k}$.
This implies that $\beta + \gamma =\alpha + \gamma \Rightarrow\ \alpha=\beta \Rightarrow\ \alpha=\beta=\gamma$ giving $m_i=m_j$.
\vspace{0.2cm}

\noindent
\textbf{\emph{Case 2:}}\\
$u_{ii}=1/m_i,\ u_{ij}=1/m_j$ in row $i$ and $u_{ji}=0,\ u_{jj}=1/m_j$ in row $j$.
This case is symmetrical to Case 1 and therefore we get that $\alpha=\beta=\gamma$ giving $m_i=m_j$.
\vspace{0.2cm}

\noindent
\textbf{\emph{Case 3:}}\\
$u_{ii}=1/m_i,\ u_{ij}=1/m_j$ in row $i$ and $u_{ji}=1/m_i,\ u_{jj}=1/m_j$ in row $j$.
Thus $\gamma\ge 2$ and therefore equation \eqref{ProofBDDTaylorTwoA} gives
\begin{linenomath*}\begin{align}
&
	2\left(\frac{m_i+m_j}{m_i m_j}\right)^k
	=
	\frac{\alpha}{m_i^{k+1}}+
	\frac{\beta}{m_j^{k+1}}+
	\gamma\left(\frac{m_i+m_j}{m_i m_j}\right)^{k+1}
\Rightarrow
\nonumber
\\
&
	2\left(\frac{\alpha+\beta+2\gamma}{(\alpha+\gamma)(\beta+\gamma)}\right)^k
	=
	\frac
	{
		\alpha(\beta+\gamma)^{k+1} + \beta(\alpha+\gamma)^{k+1} + \gamma(\alpha+\beta+2\gamma)^{k+1}
	}
	{
		[(\alpha+\gamma)(\beta+\gamma)]^{k+1}
	}
\Rightarrow
\nonumber
\\
&
	2\left( \alpha+\beta+2\gamma \right)^k
	=
	\frac
	{
		\alpha(\beta+\gamma)^{k+1} + \beta(\alpha+\gamma)^{k+1} + \gamma(\alpha+\beta+2\gamma)^{k+1}
	}
	{
		(\alpha+\gamma)(\beta+\gamma)
	}
\Rightarrow
\nonumber
\\
&
	2\left( \alpha+\beta+2\gamma \right)^k
	=
	\frac{\alpha(\beta+\gamma)^k}{\alpha+\gamma}+
	\frac{\beta(\alpha+\gamma)^k}{\beta+\gamma}+
	\frac{(\alpha\gamma+\beta\gamma+2\gamma^2)(\alpha+\beta+2\gamma)^{k} }{\alpha\beta+\alpha\gamma+\beta\gamma+\gamma^2 }
\Rightarrow
\nonumber
\\
&
	\frac{(2\alpha\beta+\alpha\gamma+\beta\gamma)(\alpha+\beta+2\gamma)^{k} }{\alpha\beta+\alpha\gamma+\beta\gamma+\gamma^2 }
	=
	\frac{\alpha(\beta+\gamma)^k}{\alpha+\gamma}+
	\frac{\beta(\alpha+\gamma)^k}{\beta+\gamma}.
\nonumber
\end{align}\end{linenomath*}
As $k\to\infty$, we get 
\mbox{ $(\alpha+\beta+2\gamma)^{k} \ne (\beta+\gamma)^k +(\alpha+\gamma)^k$ }
since $\alpha+\beta+2\gamma>\beta+\gamma,\ \alpha+\gamma$ hence we want $2\alpha\beta+\alpha\gamma+\beta\gamma=0\Rightarrow \alpha,\beta=0$ giving $m_i=m_j$.
\vspace{0.2cm}

\noindent
\textbf{\emph{Case 4:}}\\
 $u_{ii}=1/m_i,\ u_{ij}=0$ in row $i$ and $u_{ji}=0,\ u_{jj}=1/m_j$ in row $j$.
 Thus $\alpha,\beta\ge 1$ and therefore equation  \eqref{ProofBDDTaylorTwoA} gives
\begin{linenomath*}\begin{align}
&
	1/m_i^k + 1/m_j^k =
	\frac{\alpha}{m_i^{k+1}}+
	\frac{\beta}{m_j^{k+1}}+
	\gamma\left(\frac{m_i+m_j}{m_i m_j}\right)^{k+1}
\Rightarrow
\nonumber
\\
&
	\frac{1}{(\alpha+\gamma)^{k}} +  \frac{1}{(\beta+\gamma)^{k}}
	=
	\frac{\alpha}{(\alpha+\gamma)^{k+1}} + \frac{\beta}{(\beta+\gamma)^{k+1}} +
	\gamma \left(\frac{\gamma+\beta+2\gamma}{(\alpha+\gamma)(\beta+\gamma)}\right)^{k+1}	
\Rightarrow
\nonumber
\\
&
	\frac
	{
		(\beta+\gamma)^{k} + (\alpha+\gamma)^{k}
	}
	{
		[(\alpha+\gamma)(\beta+\gamma)]^{k}
	}
	=
	\frac
	{
		\alpha(\beta+\gamma)^{k+1} + \beta(\alpha+\gamma)^{k+1} + \gamma(\alpha+\beta+2\gamma)^{k+1}
	}
	{
		[(\alpha+\gamma)(\beta+\gamma)]^{k+1}
	}
\Rightarrow
\nonumber
\\
&
	(\beta+\gamma)^{k} + (\alpha+\gamma)^{k}
	=
	\frac
	{
		\alpha(\beta+\gamma)^{k+1} + \beta(\alpha+\gamma)^{k+1} + \gamma(\alpha+\beta+2\gamma)^{k+1}
	}
	{
		(\alpha+\gamma)(\beta+\gamma)
	}
\Rightarrow
\nonumber
\\
&
	(\beta+\gamma)^{k} + (\alpha+\gamma)^{k}
	=
	\frac{\alpha(\beta+\gamma)^k}{\alpha+\gamma}+
	\frac{\beta(\alpha+\gamma)^k}{\beta+\gamma}+
	\frac{\gamma(\alpha+\beta+2\gamma)^{k+1} }{\alpha\beta+\alpha\gamma+\beta\gamma+\gamma^2 }
.
\nonumber
\end{align}\end{linenomath*}
As $k\to\infty$, we get
$0 \ne (\alpha+\beta+2\gamma)^k$
since $\alpha,\beta\ge 1$ hence we require that $\gamma=0$ to get an equality. 
\vspace{0.2cm}

\noindent
\textbf{\emph{Conclusion from all the cases above}}\\
We see that $m_i\neq m_j$ is potentially possible only in Case 4.
However, $\mathbf{U}$ is strongly connected. If one connects $i$ and $j$ by a path $i=i_0, i_1, i_2, \ldots i_n=j$,
then one has $m_{i_k} = m_{i_{k+1}}$ as $i_k$ and $i_{k+1}$ must fall into Case 1, Case 2 or Case 3 above.
Thus  $m_i= m_j$.
This implies that every column of $\mathbf{U}$ has $2\le m \le N$ nonzero elements, including $u_{nn}$, that are all equal to $1/m$.
This is also true for every row of $\mathbf{U}$ because it is right stochastic by definition.

\stepcounter{steps}
\subsubsection*{Step \arabic{steps}: There exists state $S$ such that $\mathcal{C}_a=\mathcal{C}_{a'}$ for all $a,a'\in S$} 
We can define the state  $\mathcal{R}_x=\{n:u_{xn}=u_{xx}\}$ then, by definition, $x\in \mathcal{R}_x$ and $|\mathcal{R}_x|=m$ since there are $m$ nonzero elements in row $x$ of $\mathbf{U}$.
Consider the state $S=\mathcal{R}_x \setminus \{y\} $ for $y \in \mathcal{R}_x\setminus\{x\}$.
For this $S$ (as well as any other state), we have that
\begin{linenomath*}\begin{align}
	\begin{rcases}
	\text{if } n \in S \text{ then} & 1/m\\
	\text{if } n \notin S \text{ then} & 0
	\end{rcases}
	\le U(n,S) \le
	\frac{\min (m,|S|)}{m}.
\nonumber
\end{align}\end{linenomath*}
We can therefore write equation \eqref{ProofBDDTaylorSeriesCoefficients} in the form
\begin{linenomath*}\begin{align}
	\sum_{i=1}^{\min(m,|S|)} \lambda_S(i) \left( \frac{i}{m} \right)^k
	=
	\sum_{i=0}^{\min(m,|S|)} \lambda'_S(i) \left( \frac{i}{m} \right)^{k+1}
\label{ProofBDDTaylorGeneralS}
\end{align}\end{linenomath*}
where $\lambda_S(i)$ is the number of $U(n,S)$ terms equal to $i/m$ for $n\in S$ and $\lambda'_S(i)$ is the number of $U(n,S)$ terms equal to $i/m$ for $n\in \mathcal{N}$, which means that $\lambda'_S(i)\ge \lambda_S(i)$ for $i\ne 0$.
The ratio of the left-hand side and right-hand side of equation \eqref{ProofBDDTaylorGeneralS} should always be equal to one. 
Therefore, as $k\to \infty$, we require that
\begin{linenomath*}\begin{align}
	\lambda_S(i_{\max})=\lambda'_S(i_{\max})\frac{i_{\max}}{m}
\nonumber
\end{align}\end{linenomath*}
where $i_{\max}$ is the largest $i$ such that $\lambda_S(i)>0$.

We have that $i_{\max}=m-1$ in equation \eqref{ProofBDDTaylorGeneralS} because $|S|=m-1$ so $U(x,S) = (m-1)/m$.
This means that for state $S$, as $k\to \infty$, we require that
\begin{linenomath*}\begin{align}
	\lambda_{S}(m-1)=\lambda'_{S}(m-1)\frac{m-1}{m}.
\nonumber
\end{align}\end{linenomath*}
Since
	$\lambda_{S}(m-1)$
is an integer,
	$\lambda'_{S}(m-1)$
has to be a multiple of $m$ and the only possible value that satisfies this criteria is
	$\lambda'_{S}(m-1)=m$
hence
	\mbox{$\lambda_{S}(m-1)=m-1$}.

Since $\lambda'_{S}(m-1)=m$ there exist $m$ rows $j_1,j_2,\ldots,j_m$ such that $U(j_n, S)= (m - 1)/m$, that is, $u_{j_n a}=1/m \ \forall a \in S$. 
This means that $\mathcal{C}_a=\{j_1,j_2,\ldots,j_m\} \ \forall a \in S$ hence $\mathcal{C}_a=\mathcal{C}_{a'}$ for all $a,a' \in S$.


\stepcounter{steps}
\subsubsection*{Step \arabic{steps}: $m=2$ or $m=N$}
By contradiction, assume that $2<m<N$.
We can consider another state $S'=\mathcal{R}_x\setminus\{z\}$ such that $z\in \mathcal{R}_x\setminus\{x,y\}$.
We then have that $i_{\max}=m-1$ in equation \eqref{ProofBDDTaylorGeneralS} because $|S'|=m-1$ so $U(x,S') = (m-1)/m$.
As before, this means that $\mathcal{C}_a=\mathcal{C}_{a'}$ for all $a,a' \in S'$.
Since $x\in S,S'$ and $\mathcal{R}_x = S\cup S'$ we have that $\mathcal{C}_{a}=\mathcal{C}_{a'}$ for all $a,a'\in \mathcal{R}_x$.
For $2<m<N$ this implies that vertices $i\in \mathcal{R}_x$ are disconnected from $j\in \mathcal{N}\setminus \mathcal{R}_x$ and we therefore have disconnected graphs, a contradiction.
\end{proof}

\end{document}